\documentclass[reqno, a4paper]{amsart}
\usepackage{amsmath}
\usepackage{amssymb}
\usepackage{amsthm}

\usepackage[scale=0.8]{geometry}

\usepackage{natbib}
\usepackage{bibentry} 

\usepackage[english]{babel}
\usepackage[utf8]{inputenc}

\usepackage{subfig}
\usepackage{graphicx}

\usepackage[unicode]{hyperref}
\usepackage[active]{srcltx} 

\usepackage[bbgreekl]{mathbbol}
\usepackage{bm} 
\usepackage{MnSymbol} 
\usepackage{gensymb}
\usepackage{eurosym}

\usepackage{units}
\usepackage{tensor}
\usepackage{accents}

\usepackage{enumitem}

\usepackage{lineno}

\usepackage{multicol}

\usepackage{comment}

\usepackage{placeins}

\DeclareMathOperator{\divergence}{div}

\DeclareMathOperator{\Tr}{Tr}

\newcommand{\exponential}[1]{\ensuremath{{\mathrm e}^{#1}}}

\newcommand{\reference}{\mathrm{ref}}

\newcommand{\bydefinition}{\mathrm{def}}
\newcommand{\traceless}[1]{{#1}_{\delta}}

\newcommand{\diff}{\mathrm{d}}

\renewcommand{\vec}[1]{\ensuremath{\mathbf{#1}}}
\newcommand{\greekvec}[1]{\ensuremath{\boldsymbol{#1}}}
\makeatletter
\@ifpackageloaded{bm}%
{\renewcommand{\vec}[1]{\ensuremath{\bm{#1}}}%
\renewcommand{\greekvec}[1]{\ensuremath{\bm{#1}}}%
}{%
\relax
}
\makeatother
\newcommand{\tensorq}[1]{\ensuremath{\mathbb{#1}}}      
\newcommand{\tensorc}[1]{\ensuremath{\mathrm{#1}}}      

\newcommand{\transpose}[1]{#1^\top}
\newcommand{\transposei}[1]{#1^{-\top}}
\newcommand{\inverse}[1]{#1^{-1}}

\newcommand{\identity}{\ensuremath{\tensorq{I}}}

\newcommand{\cstress}{\tensorq{T}}
\newcommand{\cstressc}{\tensorc{T}}

\newcommand{\spstress}{\tensorq{S}_{\mathrm{R}}}

\newcommand{\deformation}{\greekvec{\chi}}

\newcommand{\fgrad}{\tensorq{F}}

\newcommand{\lcg}{\tensorq{B}}

\makeatletter
\@ifpackageloaded{bm}%
{%
\newcommand{\linstrain}{\bbespilon} 
}{%
\newcommand{\linstrain}{\tensorq{\varepsilon}}
}

\@ifpackageloaded{bm}%
{%
 
}{%

}

\@ifpackageloaded{bm}%
{%
\newcommand{\linstress}{\bbtau} 
}{%
\newcommand{\linstress}{\tensorq{\tau}}
}
\makeatother

\newcommand{\lstrain}{\tensorq{E}} 

\newcommand{\generictensor}{{\tensorq{A}}}

\newcommand{\gradasym}{\ensuremath{\tensorq{W}}}
\newcommand{\gradsym}{\ensuremath{\tensorq{D}}}

\newcommand{\gradvl}{\ensuremath{\tensorq{L}}}

\newcommand{\bvec}[1]{\vec{e}_{#1}} 

\newcommand{\bvecz}{\bvec{\hat{z}}}

\newcommand{\hatz}{\hat{z}}

\newcommand{\Heaviside}{H}


\newcommand{\Young}{\mathrm{E}}
\newcommand{\Poisson}{\mathrm{\nu}}

\newcommand{\ienergy}{\ensuremath{e}} 
\newcommand{\fenergy}{\ensuremath{\psi}} 
\newcommand{\entropy}{\ensuremath{\eta}} 

\newcommand{\temp}{\ensuremath{\theta}} 

\newcommand{\cheatvolref}{\ensuremath{c_{\mathrm{V}, \reference}}} 

\newcommand{\hfluxc}{\vec{j}_{q}}     

\newcommand{\pd}[2]{\ensuremath{\frac{\partial {#1}}{\partial {#2}}}}

\newcommand{\dd}[2]{\ensuremath{\frac{\diff {#1}}{\diff {#2}}}}

\newcommand{\fid}[1]{\ensuremath{\accentset{\triangledown}{#1}}}

\newcommand{\jfid}[1]{\ensuremath{\accentset{\vartriangle}{#1}}}

\newcommand{\mdif}[1]{\ensuremath{\dot{\overline{#1}}}}

\makeatletter
\@ifpackageloaded{tensor}
{

}{%

}
\makeatother

\makeatletter
\@ifpackageloaded{tensor}
{

}{%

}
\makeatother

\newcommand{\absnorm}[1]{\ensuremath{\left|#1\right|}}

\makeatletter
\@ifundefined{volume}{%
}%
{%
}
\makeatother

\newcommand{\tensortensor}[2]{\ensuremath{#1 \otimes #2}}
\makeatletter
\@ifpackageloaded{MnSymbol} 
{
\newcommand{\tensordot}[2]{\ensuremath{#1 \vdotdot #2}} 
}{%
\newcommand{\tensordot}[2]{\ensuremath{#1 : #2}} 
}
\makeatother

\newcommand{\tensorf}[1]{{\mathfrak{#1}}}

\newcommand{\rrplacer}{\kappa_R}       
\newcommand{\nplacer}{\kappa_{\mathnormal{p}(\mathnormal{t})}}  

\newcommand{\fgradrng}{\ensuremath{\tensorq{G}}} 

\newcommand{\gradvlrn}{\ensuremath{\gradvl_{\nplacer}}} 
\newcommand{\gradsymrn}{\ensuremath{\gradsym_{\nplacer}}} 

\newcommand{\lcgnc}{\ensuremath{\lcg_{\nplacer}}} 

\newcommand{\fgradnc}{\ensuremath{\fgrad_{\nplacer}}} 

\newcommand{\fgradtot}{\ensuremath{\fgrad_{\rrplacer}}} 

\newcommand{\elastic}{\mathrm{e}} 
\newcommand{\plastic}{\mathrm{p}} 

\newcommand{\fgradel}{\ensuremath{\fgrad_{\elastic}}} 
\newcommand{\fgradpl}{\ensuremath{\fgrad_{\plastic}}} 

\numberwithin{equation}{section}
\newcommand{\yield}{\mathrm{y}}
\newcommand{\thermal}{\mathrm{th}}
\renewcommand{\jfid}[1]{\ensuremath{\accentset{\medcircle}{#1}}}
\renewcommand{\tensorf}[1]{{\tensorq{#1}}}

\title[Inelastic response of solids]{A thermodynamic basis for implicit rate-type constitutive relations describing the inelastic response of solids undergoing finite deformation}

\author{David Cichra}
\address{
Faculty of Mathematics and Physics\\
Charles University\\
Sokolovsk\'a 83\\
Praha 8 -- Karl\'{\i}n\\
CZ 186\;75\\
Czech Republic
}
\email{cichra.david@gmail.com}

\author{V\'{\i}t Pr\r{u}\v{s}a}
\date{\today}
\address{
Faculty of Mathematics and Physics\\
Charles University\\
Sokolovsk\'a 83\\
Praha 8 -- Karl\'{\i}n\\
CZ 186\;75\\
Czech Republic
}
\email{prusv@karlin.mff.cuni.cz}

\thanks{V\'{\i}t Pr\r{u}\v{s}a thanks the Czech Science Foundation, grant number 18-12719S, for its support.}
\keywords{implicit constitutive relations, thermodynamics, inelastic response, plasticity}
\subjclass[2000]{%
74A15
}

\begin{document}

\begin{abstract}
Implicit rate-type constitutive relations utilizing discontinuous functions provide a novel approach to the purely phenomenological description of the inelastic response of solids undergoing finite deformation. However, this type of constitutive relations has been so far considered only in the purely mechanical setting, and the complete thermodynamic basis is largely missing. We address this issue, and we develop a thermodynamic basis for such constitutive relations. In particular, we focus on the thermodynamic basis for the classical elastic--perfectly plastic response, but the framework is flexible enough to describe another types of inelastic response as well.


\end{abstract}

\maketitle



\section{Introduction}
\label{sec:introduction}
\cite{rajagopal.kr:on*3,rajagopal.kr:elasticity} recognized that the classical concept of the Cauchy elastic material wherein the constitutive relation for a homogeneous isotropic elastic solid is given as
\begin{equation}
  \label{eq:56}
  \cstress = \tensorf{f}\left(\lcg\right),
\end{equation}
where $\cstress$ denotes the Cauchy stress tensor, $\lcg$ denotes the left Cauchy--Green tensor, and~$\tensorf{f}$ is a tensor-valued function, is overly restrictive and can be generalised as 
\begin{equation}
  \label{eq:57}
  \tensorf{h} \left( \cstress, \lcg\right) = \tensorq{0},
\end{equation}
where $\tensorf{h}$ is a tensor-valued function. It turns out that this seemingly small change can be very beneficial. For example~\cite{muliana.a.rajagopal.kr.ea:determining} and~\cite{gokulnath.c.saravanan.u:modeling} have shown that the novel constitutive framework~\eqref{eq:57} can be gainfully exploited in the modelling of the response of rubber. (See \cite{rajagopal.kr.saccomandi.g:novel} and~\cite{bustamante.r.rajagopal.kr:review} for recent review of the applications of this concept in the mechanics of solids.)

Later, \cite{rajagopal.kr.srinivasa.ar:on*5} have speculated on the use of implicit rate-type constitutive relations
\begin{equation}
  \label{eq:58}
  \tensorf{g} \left( \spstress, \dd{\spstress}{t},  \lstrain, \dd{\lstrain}{t} \right) = \tensorq{0},
\end{equation}
in the phenomenological modelling of inelastic response of solids. (Here $\spstress$ denotes the second Piola--Kirchhoff stress tensor, $\lstrain$ denotes the Green strain tensor, and $\dd{}{t}$ stands for the time derivative of the corresponding quantity.) This concept has been used by~\cite{rajagopal.kr.srinivasa.ar:inelastic} who have introduced a \emph{one-dimensional} constitutive relation 
\begin{equation}
  \label{eq:55}
  \dd{\sigma}{t}
  =
  \Young
  \left[
    1
    -
    \Heaviside
    \left(\sigma \dd{\epsilon}{t} \right)
    \Heaviside
    \left(
      \absnorm{\sigma} - \sigma_{\yield}
    \right)
  \right]
  \dd{\epsilon}{t}
  ,
\end{equation}
where $\sigma$ denotes the stress, $\epsilon$ denotes the relative deformation, $\sigma_{\yield}$ denotes the yield stress, $\Young$ denotes the Young modulus and $\Heaviside$ denotes the Heaviside step function,
\begin{equation}
  \label{eq:59}
  \Heaviside(x)
  =_{\bydefinition}
  \begin{cases}
    0, & x \leq 0, \\
    1, & x > 0.
  \end{cases}
\end{equation}
Surprisingly, the simple implicit rate-type equation~\eqref{eq:55} whose \emph{essential feature is the use of discontinuous functions} is sufficient for the modelling of the one-dimensional elastic--perfectly plastic response, and it is easy to generalise to include more complex models for plastic response, such as models for the plastic response without a sharp yield condition, see~\cite{rajagopal.kr.srinivasa.ar:inelastic} for details. (See also~\cite{wang.z.srinivasa.ar.ea:simulation} for the application of the same concept in the beam theory.)

Notable feature of the implicit rate-type equation~\eqref{eq:55} is that is allows one to model the \emph{plastic response without the need to introduce the concept of plastic strain}, which is a concept ubiquitous in classical approaches to plasticity. (See for example~\cite{steigmann.dj:primer} for a recent overview of the classical plasticity theory; exhaustive discussion and numerous references can be also found in~\cite{xiao.h.bruhns.ot.ea:elastoplasticity}.) This might be beneficial in the development of purely phenomenological models for inelastic response that deliberately make no direct reference to the evolution of the underlying microstructure, either because the microstructure evolution is of no interest or it is completely unknown or too complicated to deal with. 

While a generalisation of constitutive relations of type~\eqref{eq:55} into the fully three-dimensional setting and to finite deformations has been carried out by~\cite{rajagopal.kr.srinivasa.ar:implicit*1}, such a generalisation still works with the mechanical variables only, and it leaves the problem of establishing a thermodynamic basis for this class of models open. (Several studies, see for example~\cite{rajagopal.kr.srinivasa.ar:gibbs-potential-based}, \cite{srinivasa.ar:on}, \cite{bustamante.r.rajagopal.kr:implicit*1} or~\cite{gokulnath.c.saravanan.u.ea:representations} are focused on thermodynamics, but mainly in the context of simpler algebraic implicit constitutive relations~\eqref{eq:57}.) Consequently,\emph{ the question is whether it is possible to develop a complete thermodynamic basis that allows one to recover the mechanical models introduced by~\cite{rajagopal.kr.srinivasa.ar:implicit*1}, and that guarantees that models of this type are consistent with the first and the second law of thermodynamics}. In this short note we answer this question.

Moreover, we explicitly identify the \emph{energy storage mechanisms} and the \emph{entropy production} mechanisms, see~\cite{rajagopal.kr.srinivasa.ar:on*6}, that are related to the implicit rate-type constitutive relations for the inelastic response. Namely, we start with a well defined Helmholtz free energy and entropy production, and we derive a complete set of evolution equations, including the temperature evolution equation, for such a material.

\section{Preliminaries}
\label{sec:preliminaries}

The thermodynamic basis outlined below uses the Eulerian description, hence unlike in~\eqref{eq:58} we work with the Cauchy stress tensor $\cstress$ and the left Cauchy--Green tensor~$\lcg = _{\bydefinition} \fgrad \transpose{\fgrad}$, where $\fgrad =_{\bydefinition} \pd{\deformation}{\vec{X}}$ denotes the deformation gradient associated to the deformation~$\vec{x} = \deformation(\vec{X},t)$. Using the standard notation $\dd{}{t}$ for the material time derivative, $\gradvl$~for the velocity gradient, $\gradsym$ for the symmetric part of $\gradvl$, and $\gradasym$ for the skew-symmetric part of~$\gradvl$, we introduce the objective derivatives
\begin{subequations}
  \label{eq:60}
  \begin{align}
    \label{eq:2}
    \fid{\overline{\generictensor}} &=_{\bydefinition} \dd{\generictensor}{t} - \gradvl \generictensor - \generictensor \transpose{\gradvl}, \\
    \label{eq:3}
    \jfid{\overline{\generictensor}} &=_{\bydefinition} \dd{\generictensor}{t} - \gradasym \generictensor + \generictensor \gradasym
                                       ,
  \end{align}
\end{subequations}
where the derivative introduced in~\eqref{eq:2} is referred to as the upper convected derivative (Oldroyd derivative), while the derivative~\eqref{eq:3} is referred to as the corotational derivative (Zaremba--Jaumann derivative), see for example~\cite{marsden.je.hughes.tjr:mathematical}.

The key observation is the following. The identity $\dd{\fgrad}{t} = \gradvl \fgrad$ implies that
\begin{equation}
  \label{eq:1}
  \fid{\overline{\lcg}} = \tensorq{0}.
\end{equation}
Using~\eqref{eq:1}, we can conclude that the corotational derivative of~$\lcg$ is given by the formula
\begin{equation}
  \label{eq:4}
  \jfid{\overline{\lcg}} = \gradsym \lcg + \lcg \gradsym.
\end{equation}
Let us emphasise that this formula is a kinematical~\emph{identity} provided that $\lcg$ is the genuine left Cauchy--Green tensor. It is not a constitutive relation for a specific material. 

\section{Thermodynamics}
\label{sec:thermodynamics}

Now we are in the position to use the observation~\eqref{eq:4} in the derivation of the evolution equations for a homogeneous isotropic elastic--perfect plastic solid, which is the prime example of a solid material with an inelastic response. Let us assume that the specific Helmholtz free energy, that is the Helmholtz free energy \emph{per unit mass}, is given as
\begin{equation}
  \label{eq:14}
  \fenergy = \fenergy(\temp, \lcg_{\elastic}),
\end{equation}
where $\lcg_{\elastic}$ is a quantity that will be specified later. Our main task is to find an evolution equation for this quantity.

We note that other thermodynamics potentials might be used at this point as well. In particular, the specific Gibbs free energy might be the thermodynamic potential of choice, see~\cite{rajagopal.kr.srinivasa.ar:gibbs-potential-based} for further discussion. In fact, the Gibbs potential is the natural choice in the case of elastic bodies with constitutive relation in the form $\lcg = \tensorf{e}(\cstress)$, where $\tensorf{e}$ is a tensor valued function, see especially \cite{srinivasa.ar:on}, \cite{gokulnath.c.saravanan.u.ea:representations} and \cite{prusa.v.rajagopal.kr.ea:gibbs}, and also~\cite{rajagopal.kr.srinivasa.ar:implicit} for the case of thermoviscoelastic solids. For the sake of simplicity we however stick at Helmholtz free energy, which allows us to utilise/recover some well known formulae for Green elastic (hyperelastic) solids, and document the relation of our work to the classical nonlinear elasticity theory. A theory based on the Gibbs free energy instead of the Helmholtz free energy is definitely worth of investigation, but such a study is beyond the scope of our current contribution.

If we were dealing with an isotropic Green elastic (hyperelastic) solid, then the specific Helmholtz free energy would be given by the formula
\begin{equation}
  \label{eq:5}
  \fenergy = \fenergy(\temp, \lcg),
\end{equation}
and the evolution of $\lcg$ would be given by~\eqref{eq:4}. Since we are dealing with an elastic--perfectly plastic solid, we would like to recover~\eqref{eq:5} and~\eqref{eq:4} in the elastic regime. This motivates us to search for the evolution of~$\lcg_{\elastic}$ in the form
\begin{equation}
  \label{eq:12}
  \jfid{\overline{\lcg_{\elastic}}}
  =
  \gradsym \lcg_{\elastic}
  +
  \lcg_{\elastic} \gradsym
  +
  \tensorq{M}
  ,
\end{equation}
where the tensor $\tensorq{M}$ characterises the mismatch between the evolution of the left Cauchy--Green tensor $\lcg$ associated to the total deformation and the left Cauchy--Green tensor $\lcg_{\elastic}$ associated to the ``elastic part'' of the deformation. (Although there is no need to call it like this.) Since the evolution of~$\lcg$ is always governed by the evolution equation
\begin{equation}
  \label{eq:13}
  \jfid{\overline{\lcg}}
  =
  \gradsym \lcg
  +
  \lcg \gradsym
  ,
\end{equation}
we see that if $\tensorq{M} = \tensorq{0}$, then the evolution of $\lcg_{\elastic}$ coincides with the evolution of $\lcg$, and in this case we are dealing with the standard elastic response.

Let us now to try to identify a constitutive relation for $\tensorq{M}$ and the Cauchy stress tensor~$\cstress$. Assuming that the Helmholtz free energy is given by~\eqref{eq:14} we obtain the evolution equation for the specific entropy $\entropy$ in the form
\begin{equation}
  \label{eq:15}
  \rho \temp
  \dd{\entropy}{t}
  =
  \tensordot{\cstress}{\gradsym}
  -
  \rho
  \tensordot{
    \pd{\fenergy}{\lcg_{\elastic}}
  }
  {
    \dd{\lcg_{\elastic}}{t}
  }
  -
  \divergence \hfluxc
  ,
\end{equation}
where $\hfluxc$ denotes the heat flux vector, $\rho$ denotes the density, and $\tensordot{\tensorq{U}}{\tensorq{V}} =_{\bydefinition} \Tr \left(\tensorq{U} \transpose{\tensorq{V}} \right)$ denotes the Frobenius dot product on the space of matrices. (For details see~\cite{malek.j.prusa.v:derivation} or any standard book on continuum thermodynamics.) The definition of the corotational derivative implies that
\begin{equation}
  \label{eq:16}
  \dd{\lcg_{\elastic}}{t}
  =
  \jfid{\overline{\lcg_{\elastic}}}
  +
  \gradasym
  \lcg_{\elastic}
  -
  \lcg_{\elastic}
  \gradasym
\end{equation}\enlargethispage{1.3em}
hence the product of $\dd{\lcg_{\elastic}}{t}$ with the symmetric matrix $\pd{\fenergy}{\lcg_{\elastic}}$ can be rewritten as
\begin{equation}
  \label{eq:17}
  \rho
  \tensordot{
    \pd{\fenergy}{\lcg_{\elastic}}
  }
  {
    \dd{\lcg_{\elastic}}{t}
  }
  =
  \rho
  \tensordot{
    \pd{\fenergy}{\lcg_{\elastic}}
  }
  {
    \jfid{\overline{\lcg_{\elastic}}}
  }
  ,
\end{equation}
where we have used the fact that the matrix dot product of a symmetric matrix and a skew-symmetric matrix vanishes. Further, the assumed evolution equation~\eqref{eq:12} for $\lcg_{\elastic}$ implies that
\begin{equation}
  \label{eq:18}
  \gradsym \lcg_{\elastic}
  +
  \lcg_{\elastic} \gradsym
  =
  \jfid{\overline{\lcg_{\elastic}}}
  -
  \tensorq{M}
  .
\end{equation}
This is the \emph{Lyapunov equation} for quantity $\gradsym$, see for example~\cite{kuvc-era.v:matrix} or~\cite{silhavy.m:mechanics}. This equation can be solved explicitly provided that $\lcg_{\elastic}$ is a symmetric positive definite matrix. The solution to this equation is
\begin{equation}
  \label{eq:19}
  \gradsym
  =
  \int_{\tau=0}^{+\infty}
  \exponential{- \tau \lcg_{\elastic}}
  \left(
    \jfid{\overline{\lcg_{\elastic}}}
    -
    \tensorq{M}
  \right)
  \exponential{- \tau \lcg_{\elastic}}
  \,
  \diff \tau
  .
\end{equation}
Making use of~\eqref{eq:19} and~\eqref{eq:17} in~\eqref{eq:15} leads to the entropy evolution equation in the form
\begin{equation}
  \label{eq:20}
  \rho \temp
  \dd{\entropy}{t}
  =
  \tensordot{\cstress}
  {
    \left[
      \int_{\tau=0}^{+\infty}
      \exponential{- \tau \lcg_{\elastic}}
      \left(
        \jfid{\overline{\lcg_{\elastic}}}
        -
        \tensorq{M}
      \right)
      \exponential{- \tau \lcg_{\elastic}}
      \,
      \diff \tau
    \right]
  }
  -
  \rho
  \tensordot{
    \pd{\fenergy}{\lcg_{\elastic}}
  }
  {
    \jfid{\overline{\lcg_{\elastic}}}
  }
  -
  \divergence \hfluxc
  .
\end{equation}
Using the properties of the matrix dot product, the fact that $\pd{\fenergy}{\lcg_{\elastic}}$ commutes with $\lcg_{\elastic}$, and the identity
\begin{equation}
  \label{eq:21}
  \int_{\tau=0}^{+\infty}
  \exponential{- 2 \tau \lcg_{\elastic}}
  \,
  \diff \tau
  =
  \left[
    -\frac{1}{2}
    \inverse{\left(\lcg_{\elastic}\right)}
    \exponential{- 2 \tau \lcg_{\elastic}}
  \right]_{\tau=0}^{+\infty}
  =
  \frac{1}{2}
  \inverse{\left(\lcg_{\elastic}\right)}
  ,
\end{equation}
we can further rewrite~\eqref{eq:20} as
\begin{equation}
  \label{eq:22}
  \rho \temp
  \dd{\entropy}{t}
  =
  \tensordot{
    \left[
      \int_{\tau=0}^{+\infty}
      \exponential{- \tau \lcg_{\elastic}}
      \left(
        \cstress
        -
        2
        \rho
        \lcg_{\elastic}
        \pd{\fenergy}{\lcg_{\elastic}}
      \right)
      \exponential{- \tau \lcg_{\elastic}}
      \,
      \diff \tau
    \right]
  }
  {
    \jfid{\overline{\lcg_{\elastic}}}
  }
  \\
  -
  \tensordot{\cstress}
  {
    \left[
      \int_{\tau=0}^{+\infty}
      \exponential{- \tau \lcg_{\elastic}}
      \tensorq{M}
      \exponential{- \tau \lcg_{\elastic}}
      \,
      \diff \tau
    \right]
  }
  -
  \divergence \hfluxc
  .
\end{equation}

A brief inspection of~\eqref{eq:22} confirms that if we choose the constitutive relation for the mismatch tensor $\tensorq{M}$ as $\tensorq{M}  = \tensorq{0}$, that is if $\lcg_{\elastic} = \lcg$, and if we fix the constitutive relation for the Cauchy stress tensor as
\begin{equation}
  \label{eq:23}
  \cstress
  =_{\bydefinition}
  2
  \rho
  \lcg_{\elastic}
  \pd{\fenergy}{\lcg_{\elastic}},
\end{equation}
then the first two terms on the right-hand side of~\eqref{eq:22} vanish. This means that the material does not produce entropy due to mechanical processes, hence we are in fact dealing with the standard elastic solid, and~\eqref{eq:23} reduces to the standard representation formula for the stress in a homogeneous isotropic Green elastic (hyperelastic) solid, see for example~\cite{truesdell.c.noll.w:non-linear*1}. This is however not the case we are interested in.

We want to choose the constitutive relation for the mismatch tensor $\tensorq{M}$ in such a way that we obtain the elastic--perfect plastic behaviour. We keep the constitutive relation for the Cauchy stress tensor in the form~\eqref{eq:23}, which means that the first term on the right-hand side of~\eqref{eq:22} vanishes, and we focus on the term
\begin{equation}
  \label{eq:24}
    -
  \tensordot{\cstress}
  {
    \left[
      \int_{\tau=0}^{+\infty}
      \exponential{- \tau \lcg_{\elastic}}
      \tensorq{M}
      \exponential{- \tau \lcg_{\elastic}}
      \,
      \diff \tau
    \right]
  }
  .
\end{equation}
The choice of a constitutive relation for $\tensorq{M}$ must be done in such a way that it leads to a nonnegative entropy production, hence we need the term~\eqref{eq:24} to be nonnegative. Referring again back to the Lyapunov equation, we see that if we denote
\begin{equation}
  \label{eq:25}
  \tensorq{X}
  =_{\bydefinition}
  \int_{\tau=0}^{+\infty}
  \exponential{- \tau \lcg_{\elastic}}
  \tensorq{M}
  \exponential{- \tau \lcg_{\elastic}}
  \,
  \diff \tau
  ,
\end{equation}
then $\tensorq{M}$ solves the equation
\begin{equation}
  \label{eq:26}
  \tensorq{X} \lcg_{\elastic}
  +
  \lcg_{\elastic} \tensorq{X}
  =
  \tensorq{M}
  .
\end{equation}
Let us now choose the constitutive relation for $\tensorq{X}$ as
\begin{equation}
  \label{eq:27}
  \tensorq{X}
  =_{\bydefinition}
  -
  \Heaviside
  \left( \tensordot{\cstress}{\gradsym} \right)
  \Heaviside
  \left(
    \absnorm{\cstress} - \cstressc_{\yield}
  \right)
  \gradsym
  ,
\end{equation}
where $\Heaviside$ denotes the Heaviside step function~\eqref{eq:59} and $\cstressc_{\yield}$ is a constant.

This formula is motivated by the one-dimensional model~\eqref{eq:55}. We want $\tensorq{X}$ to be nonzero if only if the internal energy grows, that is if the material is being loaded. (Recall that the evolution equation of the specific internal energy reads $\rho\dd{\ienergy}{t} = \tensordot{\cstress}{\gradsym} - \divergence \hfluxc$.) Furthermore, we want $\tensorq{X}$ to be nonzero if only if the stress has reached a critical value. Formula~\eqref{eq:27} satisfies these requirements. We note that the function
$
\Heaviside
\left(
  \absnorm{\cstress} - \cstressc_{\yield}
\right)
$
can be replaced as needed by any reasonable yield criterion such as the von~Mises yield criterion, and the internal energy increase/decrease indicator $\Heaviside\left( \tensordot{\cstress}{\gradsym} \right)$ might be modified as needed as well. (See~\cite{rajagopal.kr.srinivasa.ar:inelastic,rajagopal.kr.srinivasa.ar:implicit*1} for some interesting modifications.)

If $\tensorq{X}$ is chosen as in~\eqref{eq:27}, then the entropy production term~\eqref{eq:24} reads
\begin{equation}
  \label{eq:29}
      -
  \tensordot{\cstress}
  {
    \left[
      \int_{\tau=0}^{+\infty}
      \exponential{- \tau \lcg_{\elastic}}
      \tensorq{M}
      \exponential{- \tau \lcg_{\elastic}}
      \,
      \diff \tau
    \right]
  }
  =
  -
  \tensordot{\cstress}{\tensorq{X}}
  =
  \Heaviside
  \left( \tensordot{\cstress}{\gradsym} \right)
  \Heaviside
  \left(
    \absnorm{\cstress} - \cstressc_{\yield}
  \right)
  \tensordot{\cstress}{\gradsym}
  \geq 0
  ,
\end{equation}
and it is nonnegative in virtue of the properties of the Heaviside function and the presence of the product $\Heaviside  \left( \tensordot{\cstress}{\gradsym} \right) \tensordot{\cstress}{\gradsym}$. This manipulation provides the rationale for the presence of $\gradsym$ in the proposed formula~\eqref{eq:27} for $\tensorq{X}$. Using~\eqref{eq:29} it is straightforward to see that if the constitutive relation for~$\tensorq{X}$ is~\eqref{eq:27}, and if the constitutive relation for $\cstress$ is~\eqref{eq:23}, then the entropy evolution equation~\eqref{eq:22} reads
\begin{equation}
  \label{eq:30}
    \rho \temp
  \dd{\entropy}{t}
  =
  \Heaviside
  \left( \tensordot{\cstress}{\gradsym} \right)
  \Heaviside
  \left(
    \absnorm{\cstress} - \cstressc_{\yield}
  \right)
  \tensordot{\cstress}{\gradsym}
  -
  \divergence \hfluxc
  ,
\end{equation}
and the nonnegativity of the entropy production is granted. (Provided that the constitutive relation for the heat flux is correctly selected; a simple choice is the standard Fourier law.) We can also note that the entropy is being produced due to the mechanical processes if and only if the mismatch tensor~$\tensorq{M}$ is nonzero, that is if the material response is not elastic. Using the standard manipulation, see for example~\cite{hron.j.milos.v.ea:on}, the evolution equation~\eqref{eq:30} can be then converted to the evolution equation for the temperature.

In order to get a complete system of governing equations, we however need to find an explicit equation for $\lcg_{\elastic}$; the quantity~$\tensorq{X}$ itself is of no direct interest. The evolution equation for $\lcg_{\elastic}$ is \eqref{eq:12}, where $\tensorq{M}$ is given in terms of $\tensorq{X}$ in virtue of~\eqref{eq:26}. Using the constitutive relation for $\tensorq{X}$, we see that
\begin{equation}
  \label{eq:31}
  \tensorq{M}
  =
  -
  \Heaviside
  \left( \tensordot{\cstress}{\gradsym} \right)
  \Heaviside
  \left(
    \absnorm{\cstress} - \cstressc_{\yield}
  \right)
  \left(
    \gradsym \lcg_{\elastic}
    +
    \lcg_{\elastic} \gradsym
  \right)
  ,
\end{equation}
which implies that the evolution equation~\eqref{eq:12} reads
\begin{equation}
  \label{eq:32}
  \jfid{\overline{\lcg_{\elastic}}}
  =
  \left[
    1
    -
    \Heaviside
    \left( \tensordot{\cstress}{\gradsym} \right)
    \Heaviside
    \left(
      \absnorm{\cstress} - \cstressc_{\yield}
    \right)
  \right]
  \left(
    \gradsym \lcg_{\elastic}
    +
    \lcg_{\elastic} \gradsym
  \right)
  .
\end{equation}

We note that if the material is not being loaded, $\tensordot{\cstress}{\gradsym} \leq 0$, or if the yield stress $\cstressc_{\yield}$ has not been reached, $\absnorm{\cstress} < \cstressc_{\yield}$, then~\eqref{eq:32} implies
\begin{equation}
  \label{eq:33}
  \jfid{\overline{\lcg_{\elastic}}}
  =
  \gradsym \lcg_{\elastic}
  +
  \lcg_{\elastic} \gradsym
  ,
\end{equation}
which means that $\lcg_{\elastic}$ satisfies the same evolution equation as the left Cauchy--Green tensor $\lcg$ for the total deformation. On the other hand, if the material is loaded $\tensordot{\cstress}{\gradsym} > 0$, and if the yield stress has been reached, $\absnorm{\cstress} = \cstressc_{\yield}$,  then~\eqref{eq:32} implies
\begin{equation}
  \label{eq:34}
  \jfid{\overline{\lcg_{\elastic}}}
  =
  \tensorq{0}
  ,
\end{equation}
which in turn implies that the Cauchy stress tensor is constant in the corotating frame.
\section{Complete system of governing equations}
\label{sec:compl-syst-govern}

Let us for simplicity assume that the specific Helmholtz free energy is given as $\fenergy\left(\temp, \lcg_{\elastic}\right) =_{\bydefinition} \fenergy_{\thermal} \left(\temp\right) + \fenergy_{\elastic} \left(\lcg_{\elastic}\right)$, where the thermal part~$\fenergy_{\thermal}$ is given by the standard formula
\begin{equation}
  \label{eq:41}
  \fenergy_{\thermal} \left(\temp\right)
  =
  -
  \cheatvolref
  \temp
  \left(
    \ln
    \left(
      \frac{\temp}{\temp_{\reference}}
    \right)
    -
    1
  \right)
  ,
\end{equation}
where $\cheatvolref$ is the specific heat at constant volume and $\temp_{\reference}$ is a given reference temperature value. The mechanical part~$\fenergy_{\elastic} \left(\lcg_{\elastic}\right)$ can be selected as needed from plethora of existing formulae for strain-energy density, see for example~\cite{mihai.la.goriely.a:how} or~\cite{destrade.m.saccomandi.g.ea:methodical}. (We recall that we work with Helmholtz free energy per unit mass, but many authors also use the Helmholtz free energy per unit volume in the reference configuration, hence minor adjustments of the formulae from the available literature might be needed.) If we further assume that the constitutive relation for the heat flux is the Fourier law $\hfluxc = - \kappa_{\thermal} \nabla \temp$, where the thermal conductivity $\kappa_{\thermal}$ is a constant, then the complete set of governing equations in the Eulerian description is the following
\begin{subequations}
  \label{eq:35}
  \begin{align}
    \label{eq:36}
    \dd{\rho}{t} + \rho \divergence \vec{v} & = 0, \\
    \label{eq:37}
    \rho \dd{\vec{v}}{t} &= \divergence \cstress + \rho \vec{b}, \\
    \label{eq:38}
    \cstress &= 2 \rho \lcg_{\elastic} \pd{\fenergy_{\elastic}}{\lcg_{\elastic}}, \\
    \label{eq:39}
    \jfid{\overline{\lcg_{\elastic}}}
    &=
      \left[
      1
      -
      \Heaviside
      \left( \tensordot{\cstress}{\gradsym} \right)
      \Heaviside
      \left(
      \absnorm{\cstress} - \cstressc_{\yield}
      \right)
      \right]
      \left(
      \gradsym \lcg_{\elastic}
      +
      \lcg_{\elastic} \gradsym
      \right)
      ,
    \\
    \label{eq:40}
    \rho
    \cheatvolref
    \dd{\temp}{t}
    &=
    \divergence \left( \kappa_{\thermal} \nabla \temp \right)
    +
    \Heaviside
    \left( \tensordot{\cstress}{\gradsym} \right)
    \Heaviside
    \left(
    \absnorm{\cstress} - \cstressc_{\yield}
    \right)
    \tensordot{\cstress}{\gradsym}
    ,
  \end{align}
\end{subequations}%
\enlargethispage{0.9em}%
where~$\vec{b}$ denotes the specific body force. This system must be solved for $\rho$, $\vec{v}$ and~$\lcg_{\elastic}$. If necessary~\eqref{eq:35} might be rewritten in the Lagrangian description.

\section{Linearization of the governing equations}
\label{sec:line-govern-equat}

Let us develop a small strain approximation of the rate-type constitutive relation~\eqref{eq:39}. The objective is to recover the one-dimensional model~\eqref{eq:55}. For the consistency with the standard von Mieses criterion, see for example~\cite{srinivasa.ar.srinivasan.sm:inelasticity}, we in~\eqref{eq:39} use $\Heaviside \left( \absnorm{\traceless{\cstress}}^2 - \kappa^2 \right)$ instead of~$\Heaviside \left(\absnorm{\cstress} - \cstressc_{\yield}\right)$, where $\traceless{\generictensor}$ denotes the deviatoric (traceless) part of the corresponding tensor, and $\kappa$ is a material parameter. 

We need a small strain approximation regarding the left Cauchy--Green tensor $\lcg$ for the total deformation, as well as its ``elastic'' counterpart $\lcg_{\elastic}$. Namely, we assume that
\begin{equation}
  \label{eq:44}
  \lcg \approx \identity + 2 \linstrain,
\end{equation}
where $\linstrain$ is the standard linearised strain tensor, and that
\begin{equation}
  \label{eq:45}
  \lcg_{\elastic} \approx \identity + 2 \linstrain_{\elastic}.
\end{equation}
Further, in the small strain approximation we assume that $\gradsym = \mdif{\linstrain}$. (Here the symbol $\mdif{\tensorq{a}}$ denotes the partial derivative of quantity $\tensorq{a}$ with respect to time.) The linearization of the constitutive relation for the Cauchy stress tensor $\cstress$ yields
$
\linstress = \lambda \left( \Tr \linstrain_{\elastic} \right) \identity + 2 \mu \linstrain_{\elastic},
$
that can be rewritten in terms of Young modulus $\Young$ and Poisson ratio $\Poisson$ as
\begin{equation}
  \label{eq:48}
  \linstrain_{\elastic} = \frac{1}{\Young} \left( \left(1 + \Poisson\right) \linstress - \Poisson \left( \Tr \linstress \right) \identity \right).
\end{equation}
Using~\eqref{eq:48} in the (linearised) evolution equation~\eqref{eq:39} allows us to completely eliminate the ``elastic'' part of the linearised strain $\linstrain_{\elastic}$ from the evolution equations,
\begin{equation}
  \label{eq:49}
  \frac{1}{\Young} \left( \left(1 + \Poisson\right) \mdif{\linstress} - \Poisson \left( \Tr \mdif{\linstress} \right) \identity \right)
  =
  \left[
    1
    -
    \Heaviside
    \left( \tensordot{\linstress}{\mdif{\linstrain}} \right)
    \Heaviside
    \left(
      \absnorm{\traceless{\linstress}}^2 - \kappa^2
    \right)
  \right]
  \mdif{\linstrain}
  .
\end{equation}
Finally, if the stress tensor has the form of the uniaxial tension in the direction of the~$z$-axis, that is if $\linstress = \sigma \tensortensor{\bvecz}{\bvecz}$, then the $\hatz\hatz$ component of~\eqref{eq:49} reads
\begin{equation}
  \label{eq:51}
  \frac{\mdif{\sigma}}{\Young}
  =
    \left[
    1
    -
    \Heaviside
    \left(\sigma\mdif{\epsilon} \right)
    \Heaviside
    \left(
      \frac{2}{3}\absnorm{\sigma}^2 - \kappa^2
    \right)
  \right]
  \mdif{\epsilon}
  ,
\end{equation}
where $\epsilon$ denotes the $\hatz\hatz$ component component~$\linstrain$. Equation~\eqref{eq:51} is the one-dimensional equation~\eqref{eq:55} provided that we set $\sigma_{\yield} =_{\bydefinition} \sqrt{\frac{3}{2}} \kappa$, which is the standard relation between the yield stress~$\sigma_{\yield}$ and the material parameter~$\kappa$, see~\cite{srinivasa.ar.srinivasan.sm:inelasticity}.

\section{Relation to some standard concepts in plasticity theory}
\label{sec:relat-some-stand}

So far, the tensorial quantity $\tensorq{M}$ has been interpreted as a quantity that characterises the mismatch between the evolution of the left Cauchy--Green tensor $\lcg$ associated to the total deformation and the left Cauchy--Green tensor $\lcg_{\elastic}$ associated to the ``elastic part'' of the deformation, see~\eqref{eq:12}. It might be of interest to elaborate on the link between $\tensorq{M}$ and other standard concepts in classical plasticity theories.

We focus on the link between $\tensorq{M}$ and the concept of \emph{evolving natural configuration}, see~\cite{srinivasa.ar:large} and~\cite{rajagopal.kr.srinivasa.ar:on*2}. According to~\cite{rajagopal.kr.srinivasa.ar:on*2} the total deformation gradient, denoted by~\cite{rajagopal.kr.srinivasa.ar:on*2} as $\fgradtot$, can be decomposed as
\begin{equation}
  \label{eq:6}
  \fgradtot = \fgradnc \fgradrng,
\end{equation}
where $\fgradrng$ is related to the inelastic response, while $\fgradnc$ is related to the elastic response. (In yet another classical approach to plasticity this \emph{formally} corresponds to the multiplicative decomposition of the total deformation gradient $\fgrad$ into the elastic and plastic part $\fgrad = \fgradel \fgradpl$. The underlying physical interpretation is however different. Note also that the same decomposition is used in~the work on viscoelastic rate-type fluids, see especially~\cite{rajagopal.kr.srinivasa.ar:thermodynamic} and~\cite{malek.j.rajagopal.kr.ea:on}.) If we introduce left Cauchy--Green tensor $\lcgnc$ associated to the elastic response as
\begin{equation}
  \label{eq:7}
  \lcgnc =_{\bydefinition} \fgradnc \transpose{\fgradnc},
\end{equation}
then the standard formula for the material time derivative of the total deformation gradient $\dd{\fgradtot}{t} = \gradvl \fgradtot$ implies that
\begin{equation}
  \label{eq:8}
  \dd{\lcgnc}{t} = \gradvl \lcgnc + \lcgnc \transpose{\gradvl} - 2 \fgradnc \gradsymrn \transpose{\fgradnc},
\end{equation}
where $\gradsymrn =_{\bydefinition} \frac{1}{2} \left( \gradvlrn + \transpose{\gradvlrn} \right)$ and $\gradvlrn =_{\bydefinition} \dd{\fgradrng}{t} \inverse{\fgradrng}$. If we use the definition of the corotational derivative, see~\eqref{eq:3}, we see that~\eqref{eq:8} can be rewritten as
\begin{equation}
  \label{eq:9}
  \jfid{\overline{\lcgnc}} = \gradsym \lcgnc + \lcgnc \gradsym - 2 \fgradnc \gradsymrn \transpose{\fgradnc}.
\end{equation}
Since we interpret $\lcg_{\elastic}$ as the left Cauchy--Green tensor $\lcg_{\elastic}$ associated to the ``elastic part'' of the deformation, which is the same interpretation as in~\cite{rajagopal.kr.srinivasa.ar:on*2}, we see that we are in position to identify $\lcgnc \equiv \lcg_{\elastic}$. Using~\eqref{eq:9} and~\eqref{eq:12} we can then make the identification
\begin{equation}
  \label{eq:10}
  \tensorq{M} = - 2 \fgradnc \gradsymrn \transpose{\fgradnc}.
\end{equation}
This gives us the \emph{sought link between $\tensorq{M}$ and quantities that arise in the approach based on the concept of natural configurations}.

Furthermore, if we use~\eqref{eq:10}, then we can rewrite the term
$
-
  \tensordot{\cstress}
  {
    \left[
      \int_{\tau=0}^{+\infty}
      \exponential{- \tau \lcg_{\elastic}}
      \tensorq{M}
      \exponential{- \tau \lcg_{\elastic}}
      \,
      \diff \tau
    \right]
  }
$
that appears on the right-hand side of the entropy evolution equation~\eqref{eq:22} as
\begin{equation}
  \label{eq:28}
  -
  \tensordot{\cstress}
  {
    \left[
      \int_{\tau=0}^{+\infty}
      \exponential{- \tau \lcg_{\elastic}}
      \tensorq{M}
      \exponential{- \tau \lcg_{\elastic}}
      \,
      \diff \tau
    \right]
  }
  =
  \tensordot{\left( \transpose{\fgradnc} \cstress \transposei{\fgradnc} \right)}{\gradsymrn}.
\end{equation}
This equality follows from equality~\eqref{eq:26}, which upon the multiplication by $\cstress \inverse{\lcg_{\elastic}}$ and the application of the trace yields
\begin{equation}
  \label{eq:42}
  2 \Tr \left(\tensorq{X} \cstress \right) = \Tr \left(\tensorq{M} \cstress \inverse{\lcg_{\elastic}}\right),
\end{equation}
where we have used the cyclic property of the trace and the fact that the Cauchy stress tensor $\cstress$ defined as
$
\cstress
=_{\bydefinition}
2
\rho
\lcg_{\elastic}
\pd{\fenergy}{\lcg_{\elastic}}
$,
see~\eqref{eq:23}, commutes with $\lcg_{\elastic}$. Next we make use of the relation~\eqref{eq:10} between $\tensorq{M}$ and $\fgradnc$ and~$\gradsymrn$, which yields
\begin{equation}
  \label{eq:43}
  \Tr \left(\tensorq{M} \cstress \inverse{\lcg_{\elastic}}\right)
  =
  \Tr
  \left(
    - 2 \fgradnc \gradsymrn \transpose{\fgradnc}
    \inverse{
      \left(
        \fgradnc
        \transpose{\fgradnc}
      \right)
    }
    \cstress
  \right)
  =
  -
  2
  \tensordot{\gradsymrn}
  {
    \left(
      \transpose{\fgradnc}
      \cstress
      \transposei{\fgradnc}
    \right)
  }
  ,
\end{equation}
and the proposition~\eqref{eq:28} follows immediately. From~\eqref{eq:30} we however know that
$
-
  \tensordot{\cstress}
  {
    \left[
      \int_{\tau=0}^{+\infty}
      \exponential{- \tau \lcg_{\elastic}}
      \tensorq{M}
      \exponential{- \tau \lcg_{\elastic}}
      \,
      \diff \tau
    \right]
  }
  $
  is the product of rate of entropy production and the absolute temperature, which is by~\cite{rajagopal.kr.srinivasa.ar:on*2} denoted as~$\xi$, hence we see that~\eqref{eq:28} is in fact equation (38) in~\cite{rajagopal.kr.srinivasa.ar:on*2}. This observation shows the link between the entropy production expressed in terms of $\tensorq{M}$ and the entropy production expressed in terms of quantities used in the approach based on the concept of natural configuration.

  Having identified a formula for $\xi$, our approach starts to significantly depart from that by~\cite{rajagopal.kr.srinivasa.ar:on*2}. While~\cite{rajagopal.kr.srinivasa.ar:on*2} specify a formula (constitutive function) for the entropy production as a function of $\fgradnc$, $\fgradrng$ and~$\gradsymrn$, where this function is one homogeneous with respect to $\gradsymrn$, and then they employ the maximum rate of entropy production criterion, we proceed differently. Since we work with $\tensorq{M}$, we do not have access to $\fgradnc$, $\fgradrng$ and~$\gradsymrn$. (In fact we \emph{do not want} to directly employ these quantities.) Consequently, our approach must rely on the use of $\cstress$, $\gradsym$ and $\lcg_{\elastic}$ only, which for that matter also distinguishes our approach from various other theories based on the strain decomposition.

  As we have show above, it turns out that this approach is feasible. If we work with quantities $\cstress$, $\gradsym$ and $\lcg_{\elastic}$, and if we are willing to use discontinuous functions as in the one-dimensional setting studied by~\cite{rajagopal.kr.srinivasa.ar:inelastic}, then we can indeed recover the desired inelastic (plastic) response.

\section{Conclusion}
\label{sec:conclusion}
We have provided a thermodynamic basis for the rate-type implicit constitutive relations that are capable of modelling the inelastic response of solids. The proposed approach leads to three-dimensional models for the inelasic response of solids undergoing \emph{finite deformations}, and it guarantees the consistency of the arising models with the first and the second law of thermodynamics, while it still inherits the conceptual simplicity of the purely mechanical theory introduced by~\cite{rajagopal.kr.srinivasa.ar:inelastic,rajagopal.kr.srinivasa.ar:implicit*1}.

The only (mathematical) difficulty in dealing with evolution equations of type~\eqref{eq:35} is the presence of the Heaviside function of the right-hand side of~\eqref{eq:39} and~\eqref{eq:40}, since it forces one to deal with differential equations with discontinuous right-hand sides, see~\cite{filippov.af:differential}. This feature is however shared with many other popular (one-dimensional) engineering models used in the description of hysteretic phenomena such as the Bouc--Wen model, see for example~\cite{ismail.m.ikhouane.f.ea:hysteresis} or \cite{charalampakis.ae.koumousis.vk:bouc-wen} or the Duhem model, see for example~\cite{visintin.a:differential}.


\bibliographystyle{chicago}
\bibliography{vit-prusa}

\end{document}